\documentclass[aps,prb,twocolumn,superscriptaddress]{revtex4-1}
\usepackage{graphicx}
\usepackage{amsmath}
\usepackage{color}
\makeatletter
\def\@dotsep{4.5}
\begin{document}
\title{Property of one-dimensional Coulomb interaction and its possible contribution to strongly correlated systems}
\author{Yongxi Zhou}
\email[Corresponding author. Email: ]
{yongxi.zhou@umontreal.ca.}
\affiliation{D$\acute{e}$partement de Chimie, Universit$\acute{e}$
de Montr$\acute{e}$al, C.P. 6128 Succursale A, Montr$\acute{e}$al,
Qu$\acute{e}$bec H3C 3J7, Canada}

\date{\today}

\begin{abstract}
The unique property of Coulomb interaction in strict one-dimensional (1D) system is revealed that the Coulomb repulsion energy of paired electrons is divergent. As consequences, electrons in 1D system can not doubly occupy the same spatial orbital and are completely localized. Numerical simulation by time dependent Hartree-Fock approximation shows this distinct property. The '0.7 anomaly' in 1D electron gas is fully explained by the property. Its possible contribution to strongly correlated systems is discussed.
\end{abstract}


\maketitle

1D electron systems often demonstrate the strongly correlated characteristics, Coulomb interaction between electrons plays an important role but its essence still keeps elusive. For instance, in 1D electron gases, there is a famous '0.7 anomaly' \cite{thomas1996possible} phenomenon remaining unclear and controversial. It's regarded that the '0.7 anomaly' should be due to spin polarization and Coulomb interaction plays a key role \cite{thomas1998interaction}, but people do not understand why Coulomb interaction can lift the spin degeneracy. There is a latent need to illustrate the essence of 1D Coulomb interaction and its affection on electron correlations.

It is well known that density functional theory (DFT) loses its exactness for strongly correlated systems. In theoretical field, Hubbard model\cite{hubbard1963electron} is still widely employed for qualitatively explaining some strong correlation phenomenons. The key conception of Hubbard theory is the in-site Coulomb repulsion between paired electrons with opposite spins occupying the same spatial orbital,
\begin{equation}\label{Hubbard}
H_{U} = U\sum_{i}n_{i\uparrow}n_{i\downarrow},\quad U = \int \frac{\phi^{2}(\mathbf{r_{1}})\phi^{2}(\mathbf{r_{2}})}
{|\mathbf{r_{1}}-\mathbf{r_{2}}|}d\mathbf{r_{1}}d\mathbf{r_{2}}.
\end{equation}
A puzzling fact then rising up is that the Coulomb interactions have already been exactly and sufficiently included in the \textit{ab initio} formalism of Hartree-Fock \cite{szabo1996modern} or DFT \cite{parr1994density} method, considering the Coulomb repulsions not only between the paired electrons but also between all the other electrons in the other orbitals (or levels). In the closed-shell case, the Coulomb part of the Hamiltonian in a single-particle equation is \cite{szabo1996modern}
\begin{equation}\label{CoulombInt}
H_{J} = 2\sum_{j}^{N/2} \int \frac{\psi_{j}^{2}(\mathbf{r_{2}})}
{|\mathbf{r_{1}}-\mathbf{r_{2}}|}d\mathbf{r_{2}},
\end{equation}
$\{\psi_{j}, j=1,2...N/2\}$ is the singular orbitals set, the Coulomb interactions from all the orbitals are already included. If expanded by the atomic basis functions $\{\phi_{\mu}\}$ to generate the matrix representation, this operator is expressed by the sum of a series of multi-center integrals \cite{szabo1996modern}
\begin{equation}\label{CoulombIntMP}
(\mu\nu|\lambda\sigma) = \int \frac{\phi_{\mu}^{*}(\mathbf{r_{1}})\phi_{\nu}(\mathbf{r_{1}})\phi_{\lambda}^{*}(\mathbf{r_{2}})
\phi_{\sigma}(\mathbf{r_{2}})}
{|\mathbf{r_{1}}-\mathbf{r_{2}}|}d\mathbf{r_{1}}d\mathbf{r_{2}},
\end{equation}
which is similar to the U in Eq. \ref{Hubbard}, but all the other three and four-center integrals are there, means an evaluation of Coulomb repulsion more sufficient than Hubbard model. Therefore, it is very controversial to say that an \textit{ab initio} computation is inclined to give an underestimated Coulomb interaction. On the other hand, when applying Hubbard model, the U can not be evaluated exactly but used only as an adjustable parameter. One possible reason for this controversial issue is that the 1D Coulomb interaction has its distinct property which can not be embodied in conventional \textit{ab initio} formalism.

Some hidden presuppositions in the conventional \textit{ab initio} formalism of Hartree-Fock, DFT or Hubbard model should be contemplated, one of them is that electrons doubly occupy orbitals. That presupposition is suspicious in 1D electron systems, the property of Coulomb interaction determined by 1D space dimensionality will be considered carefully in this article. Although literature \cite{kventsel1981thomas} proposes low dimensional Coulomb potential with a form different from $1/r$ in order to preserve the low dimensional Gauss's law, it is still assumed here that the potential keeps $1/r$ form for 1D systems, the 1D electrons in the real physical world are just considered as confined three-dimensional (3D) ones. The basic idea in this article is that the property of the 1D Coulomb integral in Eq. \ref{Hubbard} or Eq. \ref{CoulombIntMP} is different from the 3D and two-dimensional (2D) cases. The 1D space can have its unique topological property determined by its dimensionality. The parameter U, or multi-center integrals in Eq. \ref{CoulombIntMP}, should be deliberated according to the topological property. To do the two-electron (2e) integration in Eq. \ref{Hubbard}, it is common to decompose \cite{slater1960quantum} it into two steps, integration over $\mathbf{r_{1}}$ and then integration over $\mathbf{r_{2}}$, that physically means average of Coulomb energy over electron 1 and then average over electron 2. Consider the first step,
\begin{equation}\label{OneEint}
I = \int \frac{\phi^{2}(\mathbf{r_{1}})}
{|\mathbf{r_{1}}-\mathbf{r_{2}}|}d\mathbf{r_{1}}.
\end{equation}
The position at $\mathbf{r_{1}}=\mathbf{r_{2}}$ is usually regarded as a singularity, especially in numerical simulations\cite{juselius2007parallel}. However, the singularity retains only for 1D cases. For 2D and 3D cases, it is actually not a singularity. We can employ polar (2D) and spherical (3D) coordinate systems to illustrate that point. Shift the origin to $\mathbf{r_{2}}$, correspondingly replace $|\mathbf{r_{1}}-\mathbf{r_{2}}|$ by $r$ and  $\phi(\mathbf{r_{1}})$ becomes $\phi(r,\omega)$, $\omega$ is $\theta$ for 2D and $(\theta,\phi)$ for 3D, then the integral is rewritten as
\begin{equation}\label{nDint}
I = \int\frac{\phi^{2}(r,\omega_{n})}{r}\Omega_{n}r^{n-1}dr.
\end{equation}
Here $n$ is the dimensionality. There is no singularity for 2D and 3D cases $(n>1)$ because the denominator $r$ can be compensated by the integration volume element $r^{n-1}dr$. This property determines whether the 2e Coulomb integral of Eq. \ref{CoulombIntMP} is infinite or not.
There are already several works \cite{pyykko1991elements, de1986variational, fabbri1985variational} about the 2e integrals for 2D systems. Using Slater-type \cite{de1986variational, fabbri1985variational} or Gaussian-type \cite{pyykko1991elements} basis functions, analytical results for 2D Coulomb integrals can be obtained and they are finite, as expected. Numerical simulation \cite{sullivan2001time} by time-dependent Hartree-Fock method also shows the numerical Coulomb integral between two partially superposed 2D electron waves is limited. As regards in 3D cases, the formulas of 2e integrals using the Gaussian-type basis is well-built\cite{szabo1996modern, boys1950electronic} and widely used. From their deduction of these formulas we can see that the compensation by the $r^{n-1}dr$ plays a key role to make the singularity disappear and then integrals finite. The point is, that finiteness could not hold for 1D system and therefore, the 1D
2e integrals is divergent if there is any superposition of wave functions. 1D electron's behavior should be quite distinct and we need consider it from a generalized viewpoint. To my knowledge, most of 1D system modeling works only escape that singularity by using soft-Coulomb potential \cite{ham1956electronic, hausler1993interacting}, \textit{i.e.} to cut off the Coulomb potential, or replacing it by $\delta$ potential \cite{katyurin1992variation}. This simplification by a cutoff could not give results with real physical meaning, it's shown below that different cutoff value can remarkably change the electron's behavior.

Actually, in some case we can have the analytical form of 1D 2e Coulomb integral and easily show it is divergent. Consider the ground state of a 1D parabolic potential well, the spatial orbital is a Gaussian wave packet $\psi=\sqrt{C}e^{-x^{2}/2}$, suppose it is occupied by two electrons with up and down spins. The two coordinates of $x_{1}$ and $x_{2}$ actually form an infinite plane, a transform to the polar coordinate system, $x_{1}=rcos\theta$, $x_{2}=rsin\theta$ and $dx_{1}dx_{2}\Leftrightarrow rdrd\theta$, then can be applied. The Coulomb integral becomes
\begin{equation}\label{1Dint}
I=C^{2}\int\frac{e^{-x_{1}^{2}}e^{-x_{2}^{2}}}{|x_{1}-x_{2}|}dx_{1}dx_{2}
=C^{2}\sqrt{\pi}/2\int_{0}^{2\pi}\frac{d\theta}{|cos\theta-sin\theta|}.
\end{equation}
This integration is divergent while $\theta\rightarrow \pi/4$ and $\theta\rightarrow 5\pi/4$, where $|x_{1}-x_{2}|\rightarrow 0$. If the Hubbard model is applied here, that means the parameter U in Eq. \ref{Hubbard} is infinite. An infinite Coulomb energy is non-physical and does not exist. So, it is very naturally to conceive that in a completely strict 1D system, electrons cannot occupy an orbital in pair. Moreover, no superposition of wave functions of different electrons is permitted because that will result in singularity in the Coulomb integrals and divergent energy, hence 1D electrons are always localized.

To show this distinct property's effect on the behavior of 1D electrons, a numerical simulation is done to two electrons in a 1D infinite potential well. Time-dependent Hartree-Fock approximation \cite{sullivan2001time} is employed. The two electrons are assigned opposite spins in order to neglect the exchange interaction to simplify computation. Then two time-dependent single-particle equations are built as
\begin{equation}\label{oneDHFx1}
i\hbar\frac{\partial\psi_{1}(x_{1})}{\partial t} = \frac{-\hbar^{2}}{2m}\nabla_{1}^{2}\psi_{1}(x_{1}) +
\frac{e^{2}}{4\pi\epsilon_{0}}\int dx_{2}\frac{\psi_{2}^{2}(x_{2})}{|x_{1}-x_{2}|}\psi_{1}(x_{1})
\end{equation}
\begin{equation}\label{oneDHFx2}
i\hbar\frac{\partial\psi_{2}(x_{2})}{\partial t} = \frac{-\hbar^{2}}{2m}\nabla_{2}^{2}\psi_{2}(x_{2}) +
\frac{e^{2}}{4\pi\epsilon_{0}}\int dx_{1}\frac{\psi_{1}^{2}(x_{1})}{|x_{1}-x_{2}|}\psi_{2}(x_{2})
\end{equation}
The last item in each equation is the Coulomb repulsion potential from each other. The equations are discretized by the finite difference time dependent (FDTD) \cite{shibata1991absorbing} approach, the Coulomb integrals are calculated by numerical integration method. We face the problem of how to deal with the singularity in the numerical integration. The 1D space is discretized into many grid points $\{ X_{i} \}$, the singular points are those where $x_{1}=x_{2}$, \textit{i.e.} $|x_{1}-x_{2}|=0$. One choice is to ignore the singular points while doing numerical integrations but make the spatial difference $\Delta X=|X_{i}-X_{i\pm1}|$ very small. When $\Delta X\rightarrow 0$, the integral will approach to its real value. This method is tested effective for 2D and 3D numerical integrations in which a converged finite integral value very close to the real value can be obtained even when $\Delta X$ ($\Delta Y$, $\Delta Z$) is not necessary to be super small. In the 1D case, the integral is divergent when $\Delta X\rightarrow 0$, but the speed is not quite fast, that means a very fine $\Delta X$ is needed and it hugely enhances the time costs. We can make an other choice to avoid expensive time costs resulted from too fine $\Delta X$. The singular points are included, however, instead of set $|x_{1}-x_{2}|$ as 0, a finite but very small value is to make a substitution of it. In this simulation, the value is $10^{-8} \mathbf{\AA}$. This is actually a cutoff to the Coulomb repulsion potential. For numerical simulation it is reasonable, since we only want to see the effect of a large enough Coulomb repulsion. But keep in mind that only a small enough cutoff can give you the real physics picture.

The initial wave functions are set as two Gaussian wave packets defined as
\begin{equation}\label{initial}
\psi_{1,2}(x,0)= e^{-\alpha(x\pm x_{0})^{2} + ik_{1,2}x}
\end{equation}
In my simulation, $\alpha = 100\mathbf{\AA}^{-2}$, $x_{0} = 0.5\mathbf{\AA}$, $k_{1} = 100\mathbf{\AA}^{-1}$, $k_{2} = -100\mathbf{\AA}^{-1}$, they are moving to each other to generate a collision. The value of $k$ indicates a high kinetic energy, about 41keV, ensuring the collision very strong. Fig. 1 shows the collision process and the evolution of the wave functions. From $t=32000\Delta T$, the first collision starts, it's finished at about $t=70000\Delta T$. During the collision process, no superposition takes place. Electrons totally block each other and no tunneling through is permitted. As time passes, the wave packets just collide again and again and finally disperse into many chaotic peaks in the well. The dispersed waves keep localized in their original sides. They do not form a pair to share the same orbital as they are supposed to do. One may argue that in Fig. 1(f), some tiny superpositions do appear here. That is because we use a finite cutoff to approximate the infinite Coulomb repulsion and they are just numerical errors. Actually, if the cutoff value is not so tiny enough, wave packets will pass through each other. That shows that an arbitrary cutoff of the Coulomb potential may not give you the real physics picture of 1D system. One can imagine that for real Coulomb repulsion, \textit{i.e.} an infinite small cutoff, this superposition should disappear completely, that is right the real physics in an ideal 1D system.

\begin{figure}
\includegraphics[scale=0.25,angle=270]{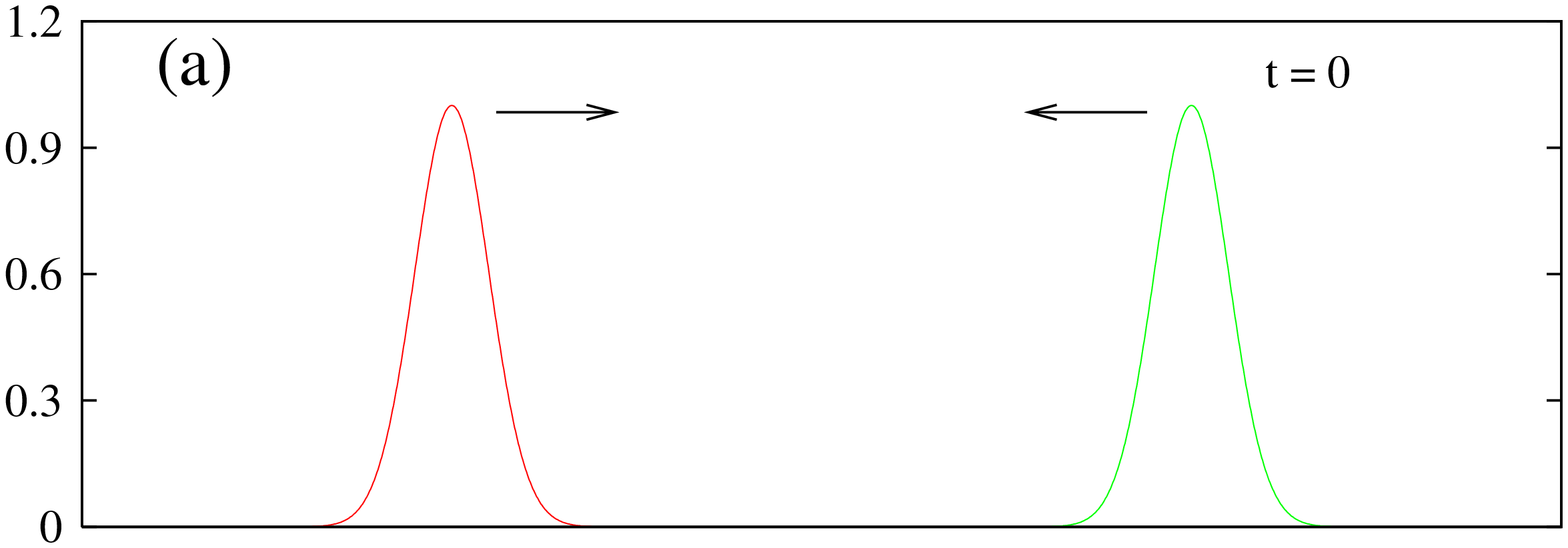}
\includegraphics[scale=0.25,angle=270]{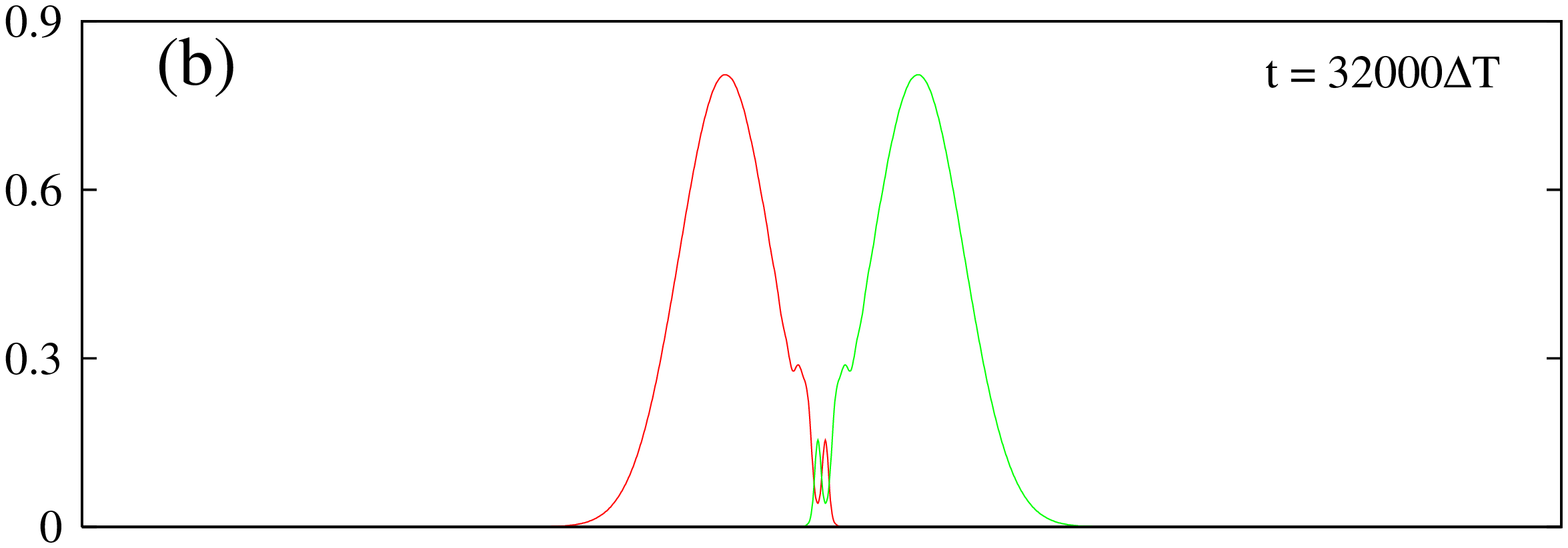}
\includegraphics[scale=0.25,angle=270]{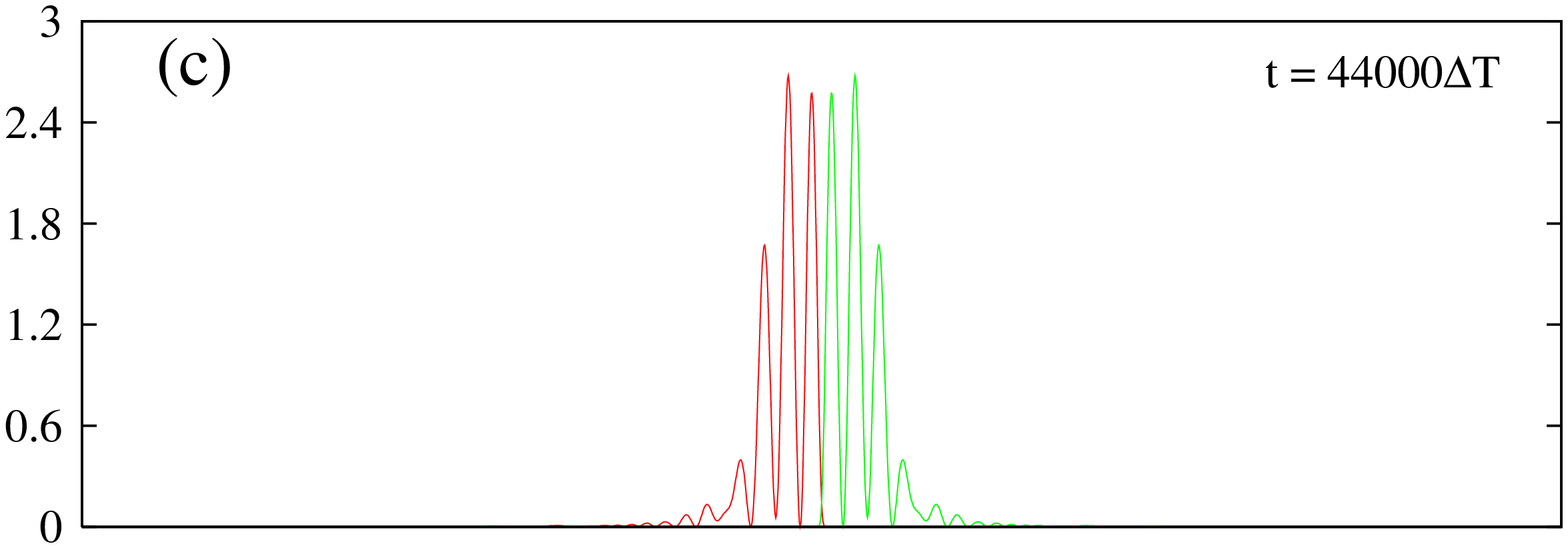}
\includegraphics[scale=0.25,angle=270]{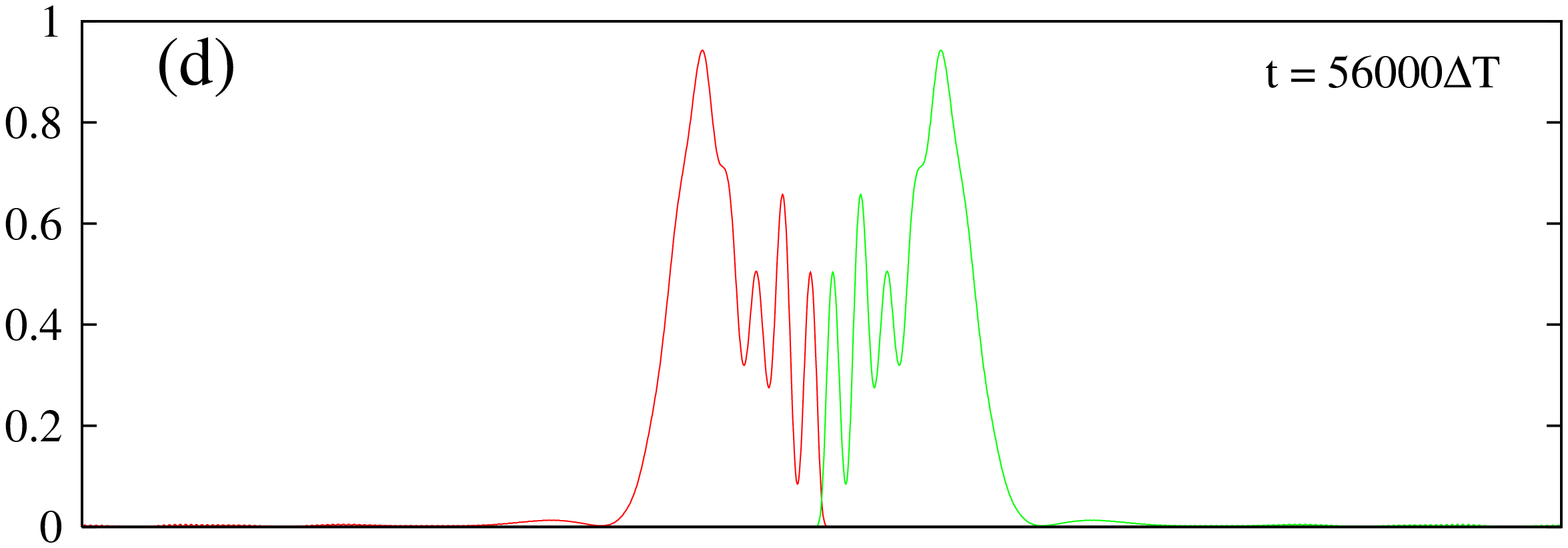}
\includegraphics[scale=0.25,angle=270]{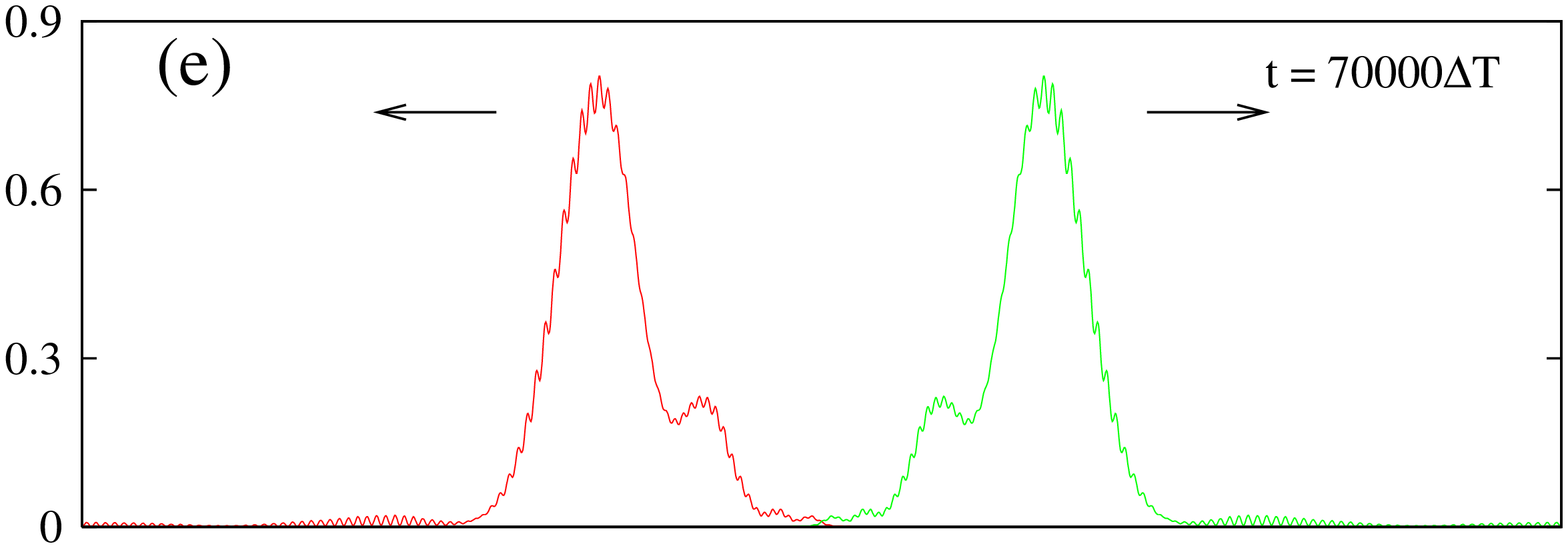}
\includegraphics[scale=0.25,angle=270]{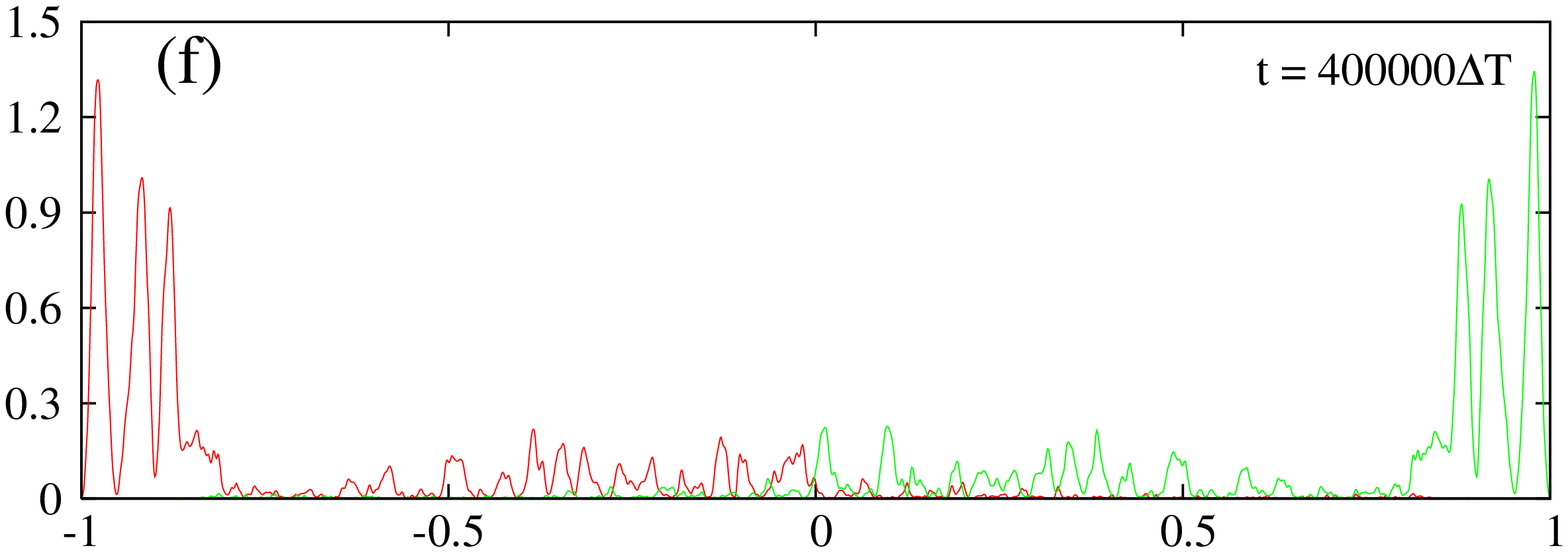}
\caption{(Color online) The collision process and evolution of the two electron wave packets in an infinite potential well simulated by FDTD method. The width of the well is 2 $\mathbf{\AA}$. Space difference is $\Delta X=0.001 \mathbf{\AA}$; temporal step is $\Delta T=10^{-8}$ femtosecond. (a) is the initial state; from (b) to (e) show the first collision process; (f) is the state after a long time of evolution, showing the dispersion and localization of the electron wave packets. }\label{collision}
\end{figure}

The conclusion that no superposition of wave functions is permitted in an ideal 1D system is extensively meaningful. Concisely, two 1D electrons with opposite spins can not occupy the same spatial orbital as they do in 2D and 3D space. More strictly, 1D electrons are totally localized. Neither electrons pass through each other nor do they overlap with each other at all. Consequently, the arrangement order of electrons is determined and electrons are distinguishable since there is no any superposition (or overlap). In other words, Pauli repulsion disappears in 1D system. Consider the exchange operator $K_{a}$
\begin{equation}\label{exchange}
K_{a}(\mathbf{r_{1}})\psi_{i}(\mathbf{r_{1}}) = \Big{[}\int d\mathbf{r_{2}}\frac{\psi_{a}^{*}(\mathbf{r_{2}})\psi_{i}(\mathbf{r_{2}})}
{|\mathbf{r_{1}}-\mathbf{r_{2}}|}\Big{]}\psi_{a}(\mathbf{r_{1}})
\end{equation}
If $\psi_{a}$ and $\psi_{i}$ have no superposition, the integration is zero, then the exchange energy is also zero. For distinguishable particles, Fermi statistics is not necessary and non-applicable. In view of the facts, we can draw a further conclusion that no band theory for 1D electrons is applicable. Many textbooks of solid physics start from the 1D periodic system to introduce the picture of band theory, but that is just not appropriate. Periodic wave functions of many electrons distributed in full 1D space must induce superpositions, which consequently result in infinite Coulomb repulsion energy. 1D electrons are always strongly correlated, free particle approximation is not applicaple. One might argue that for some particular quasi-1D systems such as carbon nanotube or atom chains, the band theory works very well. It is because these particular systems are not completely strict 1D but quasi-1D, their electron wave functions are 3D. A quasi-1D system is actually 3D one in which only the motion of electrons along one special direction is more prominent than the other two directions. Only when electrons are strongly constrained in one dimension, the 1D property can take on and its perturbation could not be ignored.

Completely strict 1D system is the extreme limitation of a constrained 3D system. To some extent, a quasi-1D, \textit{i.e.} constrained 3D, system is physically more realistic than a pure 1D one. Olszewski \cite{olszewski1982hartree} investigated such a system, interacting electron gases in a long but narrow cylindrical tube. When the radius of tube shrinks to zero, it degenerates to pure 1D. Exact analytical solutions \cite{roberts1969coulomb} of free electron in 3D cylindrical box
\begin{equation}\label{cylinder}
\phi_{n,l} = CN_{l}sin(\frac{n\pi z}{L})J_{l}(X_{l}\frac{\rho}{R})e^{il\phi}
\end{equation}
are employed as the basis functions to perform conventional Hartree-Fock calculation. Here, $L$ is the length and $R$ is the radius of the tube. The radial part $J_{l}(X_{l}\frac{\rho}{R})$ is the Bessel function, the angular part is $e^{il\phi}$. Notice that the longitudinal part $sin(\frac{n\pi z}{L})$ is some sort of plane wave basis function distributed in the whole $z$ space, then the superposition of electron waves is supposed to be there. Using these basis functions, both Olszewski \cite{olszewski1982hartree} and Roberts \cite{roberts1969coulomb} show that the Coulomb interaction integrals go to divergent while taking the limit $R\rightarrow 0$. This result is consistent with my discussion above. More interesting results are the total energies of the ground states of the gases calculated by Hartree-Fock method. As $R\rightarrow 0$, the doubly occupied case has ground state energy much higher than the singularly occupied case, and both of them are divergent while $R\rightarrow 0$. These results are not surprising. The pairing and superposition of electron waves in a pure 1D system must result in infinite Coulomb energy. Infinite energy is non-physical, then we can imagine when the $R\rightarrow 0$, \textit{i.e.} the constraint becomes strong, the real physics picture should be that the electrons must depair and localize themselves to lower the energy. If one forcefully applies conventional Hartree-Fock formalism to an ideal 1D system, only some non-physical results could be obtained because it forces electrons to be paired and delocalized.

In 1D electron gases, there is a long standing puzzle called as '0.7 anomaly' \cite{thomas1996possible} which is often considered induced by strong correlation. This phenomenon can be easily explained by my conclusions for 1D systems. In the experiment by Thomas \textit{et al.} \cite{thomas1996possible}, a gate voltage $V_{g}$ is applied to 2D electron gases (2DEG) to create 1D constrictions and control the structure width. When increasing the strength of $V_{g}$, the constriction becomes strong and the width of 2DEG becomes narrow, then finally the 2DEG is transformed to 1D gas at a large enough $V_{g}$. During this procedure, the conductance channels are closed one by one and a series of plateaus generate. Each plateau corresponds to a quantized conductance $2e^{2}/h$. The 2 is due to the doubly occupying on one channel. After the last plateau where the constriction is strong enough to squeeze the 2DEG to 1D, the anomaly value $0.7(2e^{2}/h)$ appears. This is not surprising either. As discussed above, when the constriction is very strong, electrons should access the 1D limitation and take on their 1D behaviors and depair themselves. A totally depairing makes each channel have only one electron, the quantized conductance should be $0.5(2e^{2}/h)$. If the system can not totally access the ideal 1D limitation, a partially depairing will happen and result in a conductance between $0.5(2e^{2}/h)$ and $1(2e^{2}/h)$. Denote it as $\gamma(2e^{2}/h)$. The coefficient $\gamma$ depends on the strength of constriction and the energy of conducting electrons. The value '0.7' is only one special case. Reilly \textit{et al.} \cite{reilly2001many} measured long and clean 1D wires. Very clearly the plateaus appear at $0.5(2e^{2}/h)$ while increasing the length. That's because long enough constriction makes the 1D effect remarkably prominent and the electrons almost totally depair. Thomas \textit{et al.} \cite{thomas1998interaction} suspect there is spontaneous lifting of spin degeneracy in the 1D constriction due to electron interactions. That guess is correct, however, the lifting of degeneracy should not be ascribed to magnetic mechanism but to the 1D Coulomb mechanism. This '0.7 anomaly' is a very good evidence to support my conclusions for 1D Coulomb interaction.

Up to now, theoretical researches for strongly correlated systems, even for many quasi-1D problems, are all based on these presuppositions that electrons occupy spatial orbitals in pair, they obey Fermi statistics, and unquestionably the band theory is applicable and applied. The idea that the spatial orbital be occupied by two electrons originated from early quantum theory motivated by experiments of that time and then is widely adopted in modern quantum chemistry and physics theories. The spatial parts of electron waves can be totally or partially superposed in 2D and 3D spaces. As the costs, the Coulomb repulsion energy is limitedly enhanced. However, inside some special systems, if there are very pure 1D structures or strong 1D constraints resulted from pressure, electrical or magnetic forces, these presuppositions are questionable. The property of 1D Coulomb interaction forbids any superposition of electron waves. Very possibly it is the 1D Coulomb interactions that make strong perturbations to these systems and make their behaviors so strange. Therefore, for these special systems, the formalism of conventional DFT based on 3D basis functions and 3D Coulomb integrals calculation in Eq. \ref{CoulombIntMP} cannot get correct results. Under strong 1D constraints, the electron waves are distorted, the Coulomb interaction is neither pure 3D nor pure 1D. We need develop new methods to exactly calculate these tricky Coulomb interactions. Perhaps this is also one reason why it can give us results more reasonable when we use it only as an adjustable parameter in Hubbard model.

In summary, the property of the 1D Coulomb interaction is revealed. Electrons are completely localized in 1D systems, they couldn't doubly occupy one orbital, and the Pauli repulsion does not exist. Surprisingly these properties have been ignored all along. Many theories and models for 1D electron systems just naturally adopt the assumption of doubly occupying and use the Slater determinant form of wave function to include the exchange effects. This is the first time to definitely conclude that electrons in their 1D limitation will depair and localize themselves.

I thank Matthias Ernzerhof for his hospitality, and thank Xiaogang Wang for helpful discussions. The financial supported by NSERC is gratefully acknowledged.

\bibliographystyle{apsrev4-1}
%

\end{document}